# cryo-ePDF: Overcoming Electron Beam Damage to Study the Local Atomic Structure of Amorphous ALD Aluminum Oxide Thin Films within a TEM

Ahmed M. Jasim[1], Xiaoqing He[2,3], Yangchuan Xing[1,3], Tommi A. White[2,4,*], Matthias J. Young[1,5,*]

[1]Department of Biomedical, Biological and Chemical Engineering, University of Missouri, Columbia, MO, USA

[2]Electron Microscopy Core, University of Missouri, Columbia, MO, USA

[3]Department of Mechanical & Aerospace Engineering, University of Missouri, Columbia, MO, USA

[4] Department of Biochemistry, University of Missouri, Columbia, MO, USA

[5]Department of Chemistry, University of Missouri, Columbia, MO, USA

*email: whiteto@missouri.edu or matthias.young@missouri.edu

## Abstract

Atomic layer deposition (ALD) provides uniform and conformal thin films that are of interest for a range of applications. To better understand the properties of amorphous ALD films, we need improved understanding of their local atomic structure. Previous work demonstrated measurement of how the local atomic structure of ALD-grown aluminum oxide ($AlO_x$) evolves *in operando* during growth by employing synchrotron high energy X-ray diffraction (HE-XRD). In this work, we report on efforts to employ electron diffraction pair distribution function (ePDF) measurements using more broadly available transmission electron microscope (TEM) instrumentation to study the atomic structure of amorphous ALD-$AlO_x$. We observe electron beam damage in the ALD-coated samples during ePDF at ambient temperature and successfully mitigate this beam damage using ePDF at cryogenic temperatures (cryo-ePDF). We employ cryo-ePDF and Reverse Monte Carlo (RMC) modeling to obtain structural models of ALD-$AlO_x$ coatings formed at a range of deposition temperatures from 150-332°C. From these model structures, we derive structural metrics including stoichiometry, pair distances, and coordination environments in the ALD-$AlO_x$ films as a function of deposition temperature. The structural variations we observe with growth temperature are consistent with temperature-dependent changes in the surface hydroxyl density on the growth surface. The sample preparation and cryo-ePDF procedures we report here can be used for routine measurement of ALD-grown amorphous thin films to improve our understanding of the atomic structure of these materials, establish structure-property relationships, and help accelerate the timescale for the application of ALD to address technological needs.

## 1. Introduction

Atomic layer deposition (ALD) is a well-known deposition technique for the formation of nanoscale coatings with distinct aspects such as well-controlled thickness at an atomic scale and the ability to produce conformal films on high-aspect-ratio and 3-dimensional surfaces.[1,2] These traits have made ALD an attractive technique in many applications such as catalysis,[3,4] energy storage,[5-7] water treatment,[8,9] photochemistry,[10] and beyond.[1,11] Unfortunately, ALD coatings frequently do not perform in line with expectations, and require iteration and refinement to achieve the desired outcome in a given application. Broadly, the performance of ALD coatings is known to be affected by deposition conditions such as reaction temperature, precursor family, and precursor exposure.[12-16] However, variation in the performance of ALD coatings with process conditions is not comprehensively understood due to a lack of information about the local atomic structure of ALD coatings (i.e. how atoms are arranged at a molecular level within the coatings). By improving our understanding of the atomic structure of ALD coatings, and connecting variations in atomic structure with process conditions, we expect to close the loop on the process-structure-property relationships for ALD coatings and accelerate the development timelines for ALD coatings to address specific technological needs.

ALD films are often amorphous[17] and are commonly confined to the nanoscale in at least one dimension. These aspects make it challenging to characterize the local atomic structure of ALD films. While characterization techniques such as X-ray photoelectron spectroscopy (XPS) and Fourier transform infrared spectroscopy (FTIR) are routine for the characterization of what atoms and functional groups are present in ALD coatings, they provide limited insight, (and spatial resolution), into how atoms and functional groups are assembled within the ALD coating. For crystalline phases, diffraction analysis can be used to identify atomic structure. However, for amorphous materials or structures with a low degree of order (e.g., nanoscale ALD phases), conventional diffraction analysis relying on high-intensity Bragg peaks is not viable. These difficulties are compounded in low-Z amorphous materials, such as aluminum oxide, due to weak scattering.

Despite these challenges, the atomic structure of amorphous materials can be characterized from high flux and high energy diffraction data, coupled with atomic pair distribution function (PDF) analysis.[18-20] PDF analysis is performed by taking the Fourier transform of the diffraction signal to generate a real-space representation of the diffraction data. It simultaneously examines both the diffuse and Bragg components of the diffraction pattern to reveal local and long-range order for pair distances up to 10 nm.[21] The ability to study amorphous ALD materials with PDF analysis is highlighted in previous work employing synchrotron high-energy X-ray diffraction (HE-XRD) coupled with PDF analysis to study the structure of amorphous molybdenum sulfide[22] and amorphous aluminum oxide grown by ALD on carbon nanotube (CNT) substrates,[23] as well as the structure of $InO_xH_y$ clusters[24] and $NiO_xH_y$ clusters[25] deposited within porous materials through ALD-type growth. Although HE-XRD paired with PDF analysis is a powerful tool that has been demonstrated to be successful in determining the atomic structure of amorphous ALD phases, it suffers from some shortcomings.

One key shortcoming of the use HE-XRD and PDF analysis for structural characterization of ALD coatings is the limited availability of synchrotron resources needed for HE-XRD measurements. Synchrotron resources are in high demand and require the submission of a competitive proposal for access. Synchrotrons are also geographically constrained and performing experiments requires travel to synchrotron resources and shipping any samples. These factors limit the accessibility of these resources to many researchers. Another shortcoming of HE-XRD measurements as employed in previous work to examine ALD materials,[22, 23] was the bulk nature of the measurements, requiring milligram quantities of coated powder substrate. This sample configuration limits the material systems which can be studied. For example, HE-XRD geometries preclude the study of a low-Z coating on a high-Z substrate because the substrate overwhelms the diffraction signal. The recent development of grazing-incidence HE-XRD PDF measurements[26] shows promise to help address some of these shortcomings. However, this grazing incidence approach does not allow for the examination of 3D, powder, and/or nanoscale sample geometries—preventing the study of ALD coatings on battery or catalyst particles. Developing techniques which are able to perform PDF measurements on these systems promises to help improve ALD coatings for these applications.

An alternative to HE-XRD is the use of electron diffraction (ED) within a transmission electron microscope (TEM) to perform PDF measurements, which is referred to as electron pair distribution function, or ePDF.[27] This ePDF approach addresses some of the shortcomings mentioned above for HE-XRD and provides a pathway for obtaining more localized structural information. The use of ePDF was first reported to study amorphous silicon-carbon alloys in 1986,[28] and ePDF has continued to be refined since (for a nice review of ePDF measurements see Gorelik, et al. and references contained within).[27] TEMs are relatively inexpensive (compared to synchrotron X-ray sources) and are available on many university campuses. ePDF also offers some advantages over HE-XRD/PDF, such as smaller required sample volumes, and the ability to localize the measurement area to nanoscale-spot sizes.[29-31] Relative to X-rays, electrons also have much stronger scattering interactions, allowing extremely small volumes to be probed.[32-34] Although *in situ* characterization is more challenging within a TEM relative to synchrotron X-ray diffraction studies, these advantages strongly motivate ePDF as a complementary technique to X-ray diffraction studies. Various studies have employed ePDF to examine a wide range of poorly ordered crystalline and amorphous materials. These ePDF measurements have been connected with molecular dynamics[35] and reverse Monte Carlo (RMC)[36, 37] modeling to arrive at quantitative structural models of these materials.[35, 37, 38] In recent years, efforts have continued to make ePDF more streamlined and accessible, for example by developing software tools for ePDF data analysis[39] and overcoming barriers limiting the quantitative performance of ePDF.[39-42] Together, these advances make ePDF attractive for examining the atomic structure of ALD coatings.

However, one key practical barrier to the use of ePDF to examine ALD-grown materials is the potential for damage to the sample caused by electron beam exposure.[43, 44] We note that although beam damage has been found be significant in inorganic materials,[43, 44 45-50] previous work using ePDF to study inorganic materials has largely neglected beam damage effects.[51, 52] Beam damage alters the material structure from its original state and limits the relevance of ePDF results. Electron beam damage arises because of the reaction of electrons in the electron beam with the sample, and can manifest as radiolysis, knock-on beam damage, and/or crystallization. The specific type and rate of electron beam damage varies depending on the sample composition (e.g. conductive or insulating) and electron beam conditions (electron flux, voltage and current).[53] Fortunately, the use of cryogenic temperatures has proven to be effective at mitigating beam damage during electron microscopy in many areas. In the biological science communities,[54, 55] beam damage of sensitive organic materials has been successfully mitigated using cryogenic temperatures. Cryogenic temperatures are also a critical component to the study of small molecules and biomolecules using the recently developed microED technique.[56-58] Recent studies have also begun to identify benefits of cryogenic conditions for electron microscopy studies of inorganic materials science.[59] In this context, the use of cryogenic temperatures is a natural path forward to mitigate beam damage effects during ePDF measurements to enable the study of ALD material structures.

In this work, we aim to characterize the extent beam damage effects during ePDF measurements on ALD-grown coatings and evaluate the efficacy of cryogenic temperatures to limit beam damage during ePDF measurements. We emphasize that, to our knowledge, this work represents the first study employing cryogenic temperatures during ePDF measurements. We examine amorphous ALD-grown $AlO_x$ as a test case here because it is a well-studied and widely-used ALD coating material,[12, 14, 60-63] because this material was studied in recent work using HE-XRD,[23] and because other amorphous phases of $AlO_x$ have been identified to be susceptible to beam damage.[50] We report procedures for the preparation of ALD coatings on a CNT powder substrate fixed within TEM grid support to enable rapid ePDF studies of ALD coatings. The CNT substrate we employ is a convenient material, but other substrates (e.g. silica nanospheres) are expected to perform equally well. We note that the CNT substrate employed here serves as a good model substrate for ALD growth on graphene and carbon-based materials for applications in semiconductor devices, supercapacitors, and other electrochemical devices. We demonstrate that beam damage effects are significant during ePDF measurements of ALD $AlO_x$ on CNTs at ambient temperatures, but can be successfully mitigated using cryogenic temperatures (cryo-ePDF). We then proceed to employ cryo-ePDF to study the effect of ALD deposition temperature on the atomic structure of ALD-$AlO_x$ coating layers in the range of 150-350 °C using. Results are modeled using the reverse Monte Carlo (RMC) simulations and compared with the HE-XRD measurements reported previously.[23]

## Experimental Section

### *Atomic Layer Deposition*

ALD of $AlO_x$ was performed in a custom hot-walled viscous-flow ALD reactor[64] at temperatures ranging from 150 to 400°C and maintained within <0.5°C of the setpoint using PID control. The ALD reactor was held at ~1 Torr under continuous carrier gas purge of 160 sccm argon (Ar, 99.999%, Airgas). Trimethylaluminum (TMA, 98%, Strem) and water ($H_2O$, deionized to 18.2 MΩ ionic resistance) were held at room temperature and dosed into the reactor using a virtual-valve configuration[64] with dose pressures tuned to ~200 mTorr above the Ar carrier gas pressure.

The timing sequences for one ALD growth cycle consisted of 1 s TMA dose, 10 s Ar purge, 1 s $H_2O$ dose, and 10 s Ar purge. The deposition at each temperature consisted of 100 growth cycles.

Hydroxyl-terminated multiwalled carbon nanotubes (CNTs, Nanostructured and Amorphous Materials Inc., 10-20 nm diameter) were loaded onto a TEM grid within a custom TEM grid holder for ALD deposition. This holder was constructed from 1/8" VCR components as depicted in Figure 1. When employing this holder, a TEM grid is first positioned within the VCR fitting and the fitting is sealed finger-tight (Figure 1a). Then a small quantity of CNTs powder sample is dispensed into the top tube (Figure 1b) until it visibly accumulates on the TEM grid. Once loaded with a TEM grid and CNTs, this holder is placed horizontally on a sample tray and loaded into the ALD reactor. The CNT powder on the TEM grid surface acts as the substrate for ALD depositions (Figure 1c). This holder enables the use of a small amount of CNTs and ensures rapid precursor transport to all available CNTs surfaces to produce a uniform coating on the CNTs surfaces without requiring longer dose or purge times to overcome diffusion limitations. Following deposition, the TEM grid is removed from the custom holder, and can be placed directly into a TEM sample holder for imaging and ED measurements. We note that during ALD depositions, ~2 cm square silicon wafer pieces (cut from 300 mm diameter Si wafers, Silicon Valley Microelectronics) were also placed on the sample tray both upstream and downstream of the sample holder to ensure uniform ALD growth down the length of the reactor.

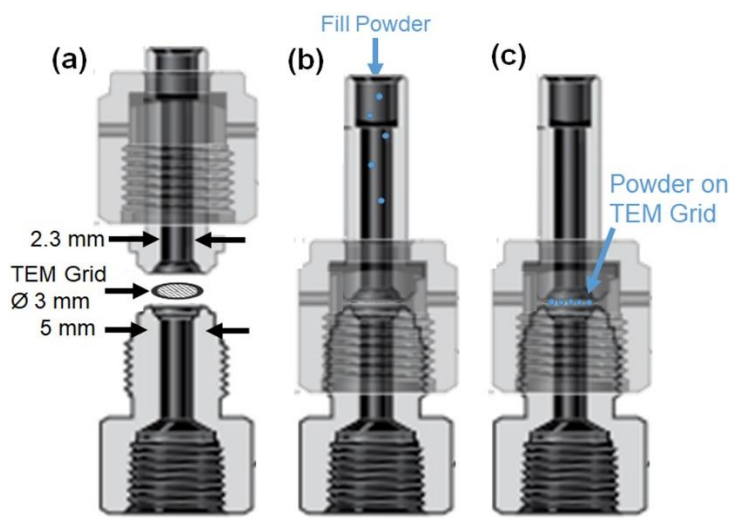

Figure 1. Schematic cartoon of a custom TEM grid holder for ALD deposition including (a) the positioning of a TEM grid within the 1/8" VCR fitting, (b) CNTs (depicted in blue) powder loading

procedure by filling through the VCR tube to settle on the TEM grid, and (c) assembled holder for TEM grid where the CNTs are resting on the TEM grid inside sample holder.

*Transmission Electron Microscopy (TEM)*

Initial TEM imaging to confirm deposition and measure film thickness were performed at room temperature on a JEOL JEM-1400 TEM equipped with Lanthanum-hexaboride (LaB$_6$) filament at an acceleration voltage of 80 kV and images collected on a Gatan Ultrascan 1000 CCD camera. After briefly confirming film thickness at 80 kV at room temperature, the samples were removed and subsequent measurements on different CNTs (to avoid any damage imparted by imaging at 80 kV) were performed using the FEI Tecnai F30 Twin TEM (FEI Co, Hillsboro, OR, USA) equipped with Gatan Ultrascan 4000 charge coupled device (CCD) with a U-type coating for ultrasensitivity. Bright-field imaging was performed at 200 kV. For cryogenic temperature measurements, TEM grids were loaded into a side-entry cryoholder (Gatan 626, Gatan Inc Pleasanton CA USA) and cooled with liquid nitrogen to cryogenic temperatures (~77 K), before loading the side-entry cryoholder into the FEI Tecnai F30 Twin TEM.

For all ED and cryo-ePDF measurements, a spot size setting of 8 was used, with a 50 µm condenser 2 aperture setting to produce a highly localized parallel beam around 200 nm in diameter, as visualized in brightfield TEM imaging mode. These settings enabled acquisition of an ED from a single CNT without the need for a selected area aperture. Unless otherwise indicated, each ED measurement was performed on an individual CNT with a ~200 nm beam diameter and 10 s electron beam exposure (measurement duration). Diffraction mode was enabled on the TEM to collect ED using a nominal camera length of 100 mm. This nominal camera length (100 mm) was calibrated for PDF analysis using the ED from a [110] silicon TEM sample. This calibration data was obtained under the same experimental conditions used in the collection of the ED from the CNTs. We performed various experiments examining the impact of lens hysteresis on diffraction measurement error, and identified a maximum error of 0.75% arising from lens hysteresis using the TEM employed in this work. To estimate the electron flux during diffraction measurements, we measured the counts on the CCD in brightfield TEM mode under the same settings as used for diffraction measurements. As this image (1024 x 1024 pixels) contains the various illumination levels due to the beam stop and Fresnel fringes from the edge of the parallel beam, a fully illuminated reduced region (128 × 128 pixels) was used to calculate the flux observed during a 0.05 s exposure. We then converted these counts using the CCD's conversion efficiency factor (26 counts/electron at 200 kV) and corrected for the scaled pixel size (1.9 Å/pixel) at the magnification used. This yielded a flux of 7.28 e$^-$/Å$^2$/s. We emphasize that the ED measurement itself involves electron beam exposure, and the duration of one measurement is 10 s. We note that for ePDF diffraction measurements, it is critical to calibrate the TEM camera length (sample-to-detector distance) and quantify lens hysteresis effects (and normalize to remove them if needed) to produce accurate PDFs.

**Pair Distribution Function Analysis and Stochastic Structural Modelling**

PDF analysis was performed using a combination of GSAS-II,[65] SUePDF,[39] and PDFgetX3[66] software packages. We note that GSAS-II and PDFgetX3 are designed primarily for synchrotron X-ray analysis, and SUePDF is designed primarily for ePDF analysis. However, considering the common elastic-scattering physics between X-ray and electron diffraction and linear scaling of both X-ray and electron scattering factors with atomic mass at high radiation energies, either software package should be viable for processing ePDF data. Indeed, we found that PDFs generated from the same input diffraction data were consistent between these software packages. As such, packages were employed as needed to take advantage of convenient features available in each package to improve the analyses we performed.

First, we used GSAS-II to import the 2-D ED pattern and process it into 1-D plot of diffraction intensity (I) vs momentum transfer (Q). We employed the De Broglie wavelength of the electron beam including relativistic effects (2.508 pm at 200 kV accelerating voltage),[67] and calibrated the sample-to-detector distance (i.e. camera length) using a calibration measurement on a crystalline silicon calibrant. We then manually aligned the beam center of the diffraction patterns by locating the center of the most prominent diffraction ring (i.e. at Q = 1.8 Å$^{-1}$ for CNTs),[68] and integrated the full 360° of diffraction data (rather than an individual line scan).

Upon generating I vs. Q data using GSAS-II for the bare CNTs and ALD-coated CNTs samples, we then established a method to subtract CNT diffraction signal from the ALD-coated CNT samples as depicted in Figure 2. For this, we first employed SUePDF to remove the smooth diffuse background present in each ED pattern (Figure 2a) arising from the direct-beam periphery, as well as inelastic and incoherent multiple scattering.[39] This produced background subtracted diffraction data as depicted in Figure 2b. We note that the smooth background varies depending on the specific material and sampling volume (i.e. diffraction path length) and needs to be removed before taking the difference between two diffraction patterns. After the smooth background was removed in SUePDF, we then manually scaled the CNT intensity as depicted in Figure 2c and subtracted the CNT diffraction component from each $AlO_x$-CNTs diffraction pattern to isolate the diffraction data from the $AlO_x$ coating as depicted in Figure 2d. We then processed this $AlO_x$ diffraction data up to a maximum Q value ($Q_{max}$) of 22 Å$^{-1}$ into PDFs using PDFgetX3. We note that for calculation of the PDFs we report below, we employed a CNT background scaled by a factor of 0.33 to remove the contributions from the CNT substrate without over-subtracting. This scaling factor produced a smooth diffraction pattern after subtraction, as depicted in Figure 2d. Both smaller and larger scaling factors down to 0.00 and up to 1.00 were examined as well. Smaller scaling factors were found to under-subtract the CNT feature at Q = 5 Å$^{-1}$ and larger scaling factors were found to produce a nonphysical negative scattering intensity at Q =1.7 Å$^{-1}$ and Q= 3.0 Å$^{-1}$ in the subtracted pattern, both resulting in CNT artifacts in the resulting PDFs. We note that a full analysis of the PDFs was also performed using a scaling factor of 1.00 on the CNT background (not shown), which yielded an average difference of 1.6% for the structural metrics reported below for ALD-grown $AlO_x$.

We note that for processing of the 2D electron diffraction patterns into PDFs, it is necessary to account for relativistic effects in calculating the De Broglie wavelength of diffracting electrons. Error can also be introduced if the beam-stop and any related artifacts are not properly masked out of the 2D diffraction data before integration, or if the beam center is not correctly positioned within the 2D diffraction pattern during integration. Also note that because of drift in the electron beam focusing lenses, the electron beam center position will vary between diffraction images and must be selected for each diffraction pattern. Drift in the electron beam can also potentially introduce error if the electron beam does not impinge perpendicular to the plane of the imaging detector surface, resulting in ellipsoidal diffraction rings. However, this effect, if present, can be measured and accounted for during data processing.

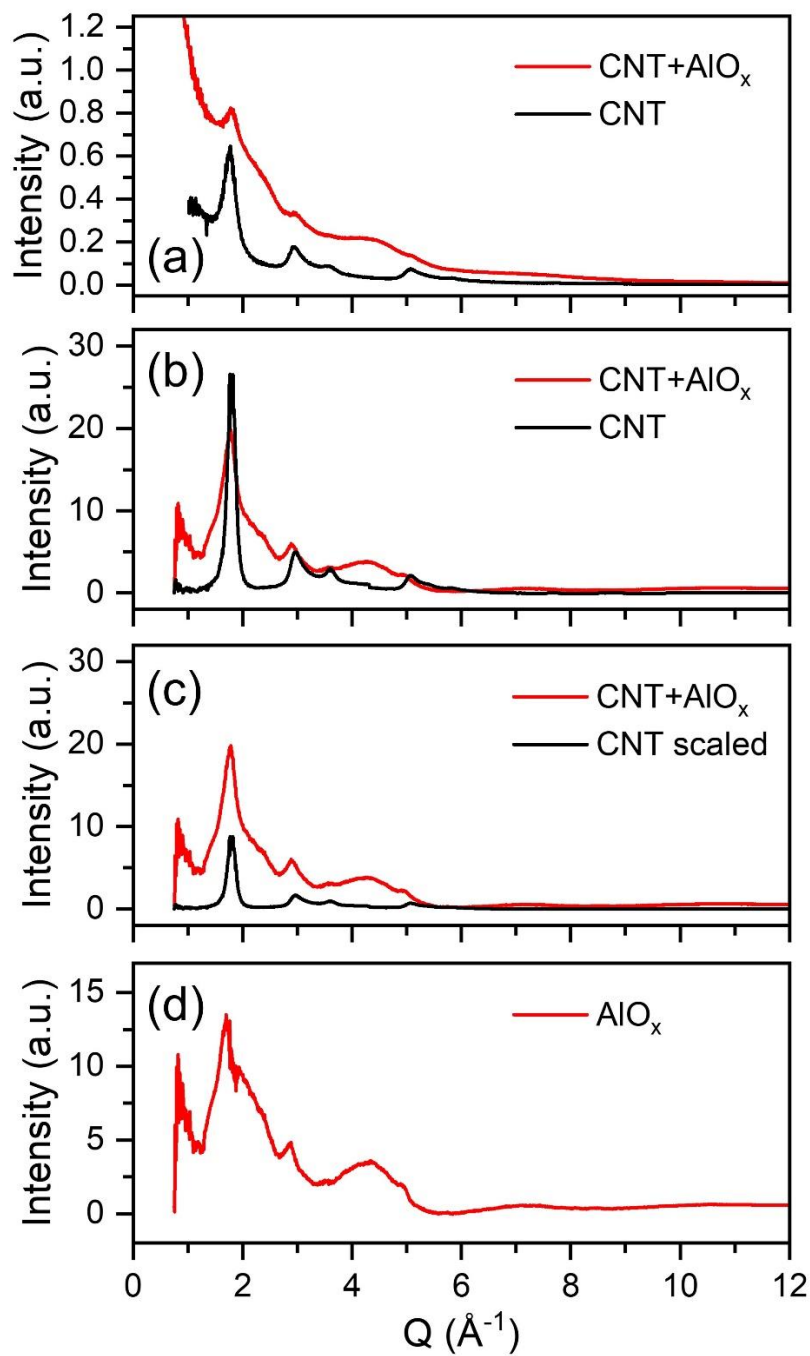

Figure 2. Example of methods for removing (a) CNT signal from ALD-coated CNT sample by (b) subtraction of diffuse background, (c) scaling of CNT diffraction intensity to match CNT signal in ALD-coated sample, and (d) subtraction of scaled CNT data from ALD-coated sample.

Stochastic reverse Monte Carlo (RMC) structural modeling was performed on the PDFs generated from $AlO_x$ using the fullrmc software package[69] to generate structural models of the ALD $AlO_x$ which were consistent with the experimental ePDF data. Simulation boxes for stochastic modeling were initiated as crystalline θ-Al$_2$O$_3$ with ≥4 nm on each edge (7840 total atoms). The structure was perturbed using > 6×10$^6$ translational and atom removal steps. The atom removal steps employed the "AtomsRemoveGenerator" implemented in fullrmc, which enables the prediction of Al:O stoichiometry. Statistical analysis of the resulting structural models was performed using ISAACS software package[70] to reveal structural information such as material composition, bonding, and atomic coordination environments.

**Spectroscopic ellipsometry**

Spectroscopic ellipsometry (SE) was performed on flat Si pieces using an alpha-SE spectroscopic ellipsometer (J.A. Woollam) at an incident angle of 65° and wavelengths from 380 to 900 nm. SE modeling was performed within the CompleteEASE software package using a Cauchy model[71, 72] for the ALD AlO$_x$ layer, where the modeled film thickness was allowed to vary for each sample. The Cauchy model was of the form $n(\lambda) = A + B/\lambda^2 + C/\lambda^4$, where n is the refractive index, λ is the wavelength, and A, B, and C are fitted constants. The constants A, B, and C were forced to be consistent for all samples and the values of the coefficients were allowed to vary to minimize the modeling error overall for the full sample set, yielding a refractive index of 1.70 at a wavelength of 580 nm, in close agreement with previous reports.[73, 74] We note that the optical properties of ALD-grown AlO$_x$ are known to vary with deposition temperature.[15] Allowing the optical properties of ALD AlO$_x$ to vary depending on deposition temperature produced a 1.3% difference in the modeled thickness values on average.

## Results and Discussion

Figure 3a-d displays TEM images showing the uniform and conformal coatings of the ALD AlO$_x$ on the CNTs. We note that the use of OH-terminated CNTs here enables facile nucleation without the use of nucleation procedures (e.g. NO$_2$/TMA) employed in previous work.[75-77] Each film was deposited using 100 ALD cycles and the coatings were deposited at a range of deposition temperatures from 150 °C to 400 °C. Figure 3e shows the coating thickness versus deposition temperature. The thickness decreases with increasing deposition temperature as measured on the CNT substrates using TEM imaging and on silicon wafer pieces using SE, consistent with previous literature reports.[15, 78, 79] This decrease in growth rate with increasing temperature is expected to arise from a decrease in surface hydroxyl density with increasing temperature,[15, 64, 80-84] and is consistent with the description of an "ALD window" in temperature arising from thermochemical effects.[85-87] The different thicknesses between SE and TEM are expected to arise from differences in nucleation between the two substrates, and difficulty discerning the exact location of the $AlO_x$/CNT interface in the TEM images.

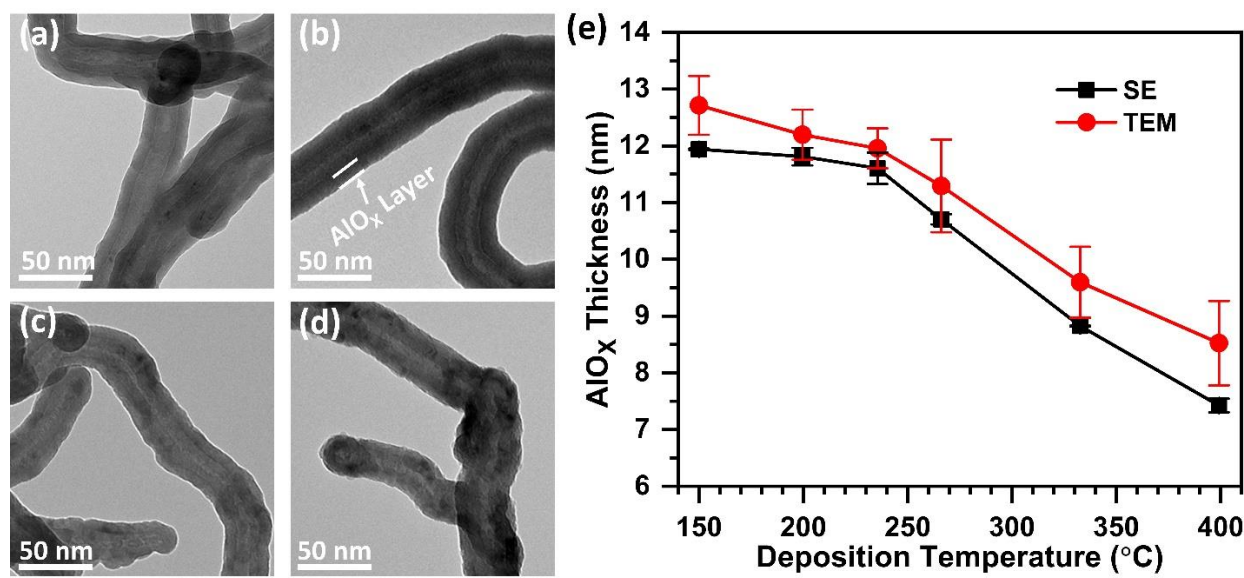

Figure 3. Brightfield TEM images acquired at ambient temperature with 80 kV accelerating voltage on JEOL JEM 1400 TEM of $AlO_x$ coating layer on CNTs deposited using 100 ALD cycles of TMA/$H_2O$ at (a) at 200 °C, (b) 267 °C, (c) 332 °C, (d) 400 °C, along with (e) plot of $AlO_x$ thickness versus the ALD-deposition temperature measured by TEM (red circles) and SE (black squares). The error bars in the SE data represent standard deviation in measured thickness for two Si samples included in the ALD reactor during deposition, and error bars in TEM data represent standard deviation in measured thickness from measurements in three different regions of TEM images.

As described in the introduction, a key barrier to ePDF analysis of ALD materials is the potential for electron beam damage. In Figure 4, we present TEM micrographs which highlight some examples of beam damage effects on ALD $AlO_x$. The sample in Figure 4a is the same sample depicted in Figure 3d after extended (~45 s) electron beam exposure at an accelerating voltage of 80 kV. We note that an 80 kV accelerating voltage is not ideal for ePDF measurement, but we report the beam damage effect we observed at this accelerating voltage to highlight the range of beam damage effects which may be observed during ePDF measurements. During extended exposure at 80 kV, we observed the formation of spherical beads on the surface of the $AlO_x$-coated CNTs. At this 80 kV accelerating voltage, the beam damage is expected to arise mostly from radiolytic decomposition,[88, 89] analogous to radiolysis effects observed in silicates.[45-48] In such a process, the incident electron beam causes electronic excitation and cleaves Al-O bonds, forming O-O defects[46, 89] and surface radicals.[90] This leads to oxygen migration to the surface to form O-O clusters and $O_2$ gas bubbles[47, 48, 89] and results in metal ion reduction and structure deformation.[91] (We note that the beads observed in Figure 4a may also arise from the degradation of the underlying CNT substrate to generate hydrogen gas bubbles, which has been reported in previous work on viruses.[92, 93]) We expect that the spheres on the surface of the $AlO_x$ in Figure 4a are formed by one or more radiolytic processes, and are either $AlO_x$ shells with voids formed by the release of $O_2$ gas bubbles[48, 89] or hydrogen gas bubbles,[94, 95] or are beads of reduced $AlO_x$/metallic Al. Although it may be possible to overcome the radiolysis effects observed in Figure 4a using cryogenic temperatures, the low (80 kV) accelerating voltage employed here limits the Q-range for ePDF measurements, thus we focus on beam damage effects at higher accelerating voltage in the following.

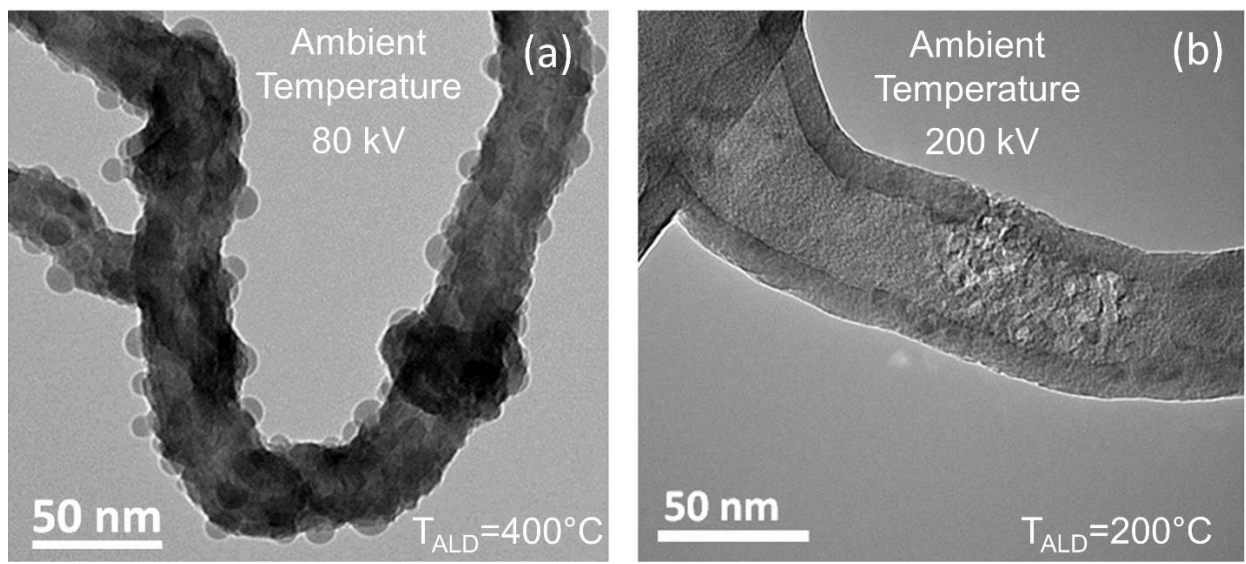

Figure 4. Brightfield TEM micrographs from two different TEMs highlighting beam damage effects observed at ambient temperature on ALD-grown $AlO_x$ including (a) radiolysis at low accelerating voltage (80 kV) in the JEOL JEM 1400 TEM (b) local crystallization observed at high accelerating voltage (200 kV) in the FEI Tecnai F30 Twin TEM, the $AlO_x$ deposition temperature is indicated in the lower right corner of each panel.

At higher accelerating voltage (200 kV) in Figure 4b, we observe a beam damage effect which is distinct from the radiolysis effect observed at the 80 kV accelerating voltage in Figure 4a. We note that the beam was localized in a smaller ~200 nm diameter area for ePDF measurement prior to taking the image in Figure 4b, and the area where the beam was focused is the same area where the damage is observed in Figure 4b. At this higher accelerating voltage, we observe the formation of nanoscale crystallite clusters. However, we do not observe expansion of the material or the emergence of spherical formations as we did at an 80 kV accelerating voltage in Figure 4a. Prior work has reported local crystallization of amorphous materials with a focused high-energy electron beam in line with our observations.[96] While some researchers have attributed the crystallization to a thermal effect,[97] most concluded that the temperature increase from a localized electron beam is minimal,[96, 98] and attribute the crystallization to electronic excitations from the incident beam.[49, 99] Electronic excitations are thought to cause defects within the amorphous phase to cluster, leading to the formation of crystalline domains.[49] Because there is a kinetic barrier associated with these structural reorganizations, crystallization is expected to slow at cryogenic temperatures, where the crystallization rate depends on the ratio of the defect decay rate to the rate of bombardment.[49] Assuming a kinetic barrier of ~0.5 eV, a reduction in temperature from 298 K to 77 K is predicted to produce a decrease in the damage rate by more than 20 orders of magnitude based on this prior work.[49]

Based on the predicted decrease in the rate of crystallization under cryogenic conditions at high accelerating voltages (as well as the success of cryogenic temperatures in mitigating beam damage in other fields), we employed a cryogenic TEM holder cooled using liquid nitrogen (~77 K) during ED. However, before measuring ALD-coated CNTs, we first evaluated the effect of cryogenic temperatures on beam damage effects in CNTs without an ALD coating. These experiments served to help us interpret the diffraction data obtained from ALD-coated CNTs below. Presented in Figure 5a and 5b are 2D diffraction patterns, each for a single bare CNT at a 200 kV accelerating voltage, measured in ambient and cryogenic temperature conditions, respectively. These measurements were each performed on individual CNTs with a ~200 nm beam diameter and 10 s electron beam exposure. We note that different colors in Figure 5a vs. Figure 5b arise because of differences in diffraction intensity leading to differences in the heat map depictions. The dominant features in each diffraction pattern are equivalent (as evident in Figure 5c, described below).

Interestingly, we observe smooth diffraction rings for diffraction on a single CNT under cryogenic temperatures as shown in Figure 5b. This is surprising because crystalline CNTs would be expected to generate a textured diffraction pattern. The uniform-intensity diffraction rings (lacking texturing) suggest that the multiwalled, hydroxyl-terminated CNTs we employ as our substrates lack structural ordering. We observe a similar lack of texturing for ALD-coated CNTs (*vide infra*). The chemically functionalized -OH terminated CNTs were used as-received from the supplier and are not graphitized. We expect that the chemical functionalization process performed by the manufacturer leads to the formation of CNTs with poor structural order in line with previous observations.[100] The disordered character of this CNT substrate is helpful to us here because the uniform diffraction intensity and lack of texturing allows for subtraction of the ED signal from the CNT substrate from a sample comprised of CNTs with an ALD coating (as demonstrated above in Figure 2). This enables us to study the atomic structure of the $AlO_x$ coating without contribution from the substrate. We note that the background subtraction procedure we report in Figure 2 is expected to only be successful if the diffraction signal from the substrate is not textured.

We observe some texturing in the ED performed on CNTs under *ambient* temperatures in Figure 5a, suggesting that beam damage at ambient temperatures leads to crystallization of the CNT in line with previous observations.[101] Indeed, in Figure 5c, we plot the integrated diffraction intensity vs. Q, and we observe that the sample measured under ambient temperature has more pronounced diffraction peaks compared to the sample measured under cryogenic temperature (note the upward shift in the ambient trace on a *LOG*-scale, indicating an increase in peak intensity). In Figure 5c we also include vertical dashed lines at dominant peak locations expected for CNTs (~1.8 $Å^{-1}$)[68] and aluminum oxide (~4.2 $Å^{-1}$).[102] Importantly, no diffraction peaks are observed around 4.2 $Å^{-1}$ in the bare CNT sample, allowing for interpretation of the beam damage in $AlO_x$ and CNTs separately in ALD coated CNT samples (*vide infra*). Overall, these data suggest that the use of cryogenic temperature may slow damage in CNTs sufficiently to allow for analysis of the $AlO_x$ coating on top of the CNTs.

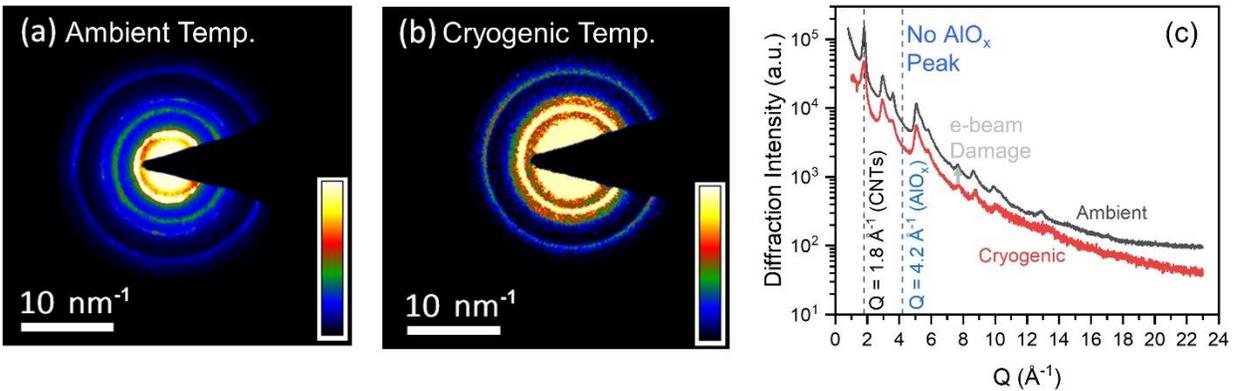

Figure 5. 2D diffraction patterns for ED of bare, uncoated CNTs under (a) ambient and (b) cryogenic temperature measurement conditions from a 10 s electron diffraction exposure at an accelerating voltage of 200 kV. The patterns have been colored using a heat map representing CCD electron counts with inset scalebars depicting relative diffraction intensity and reciprocal distance. (c) Total scattering intensity in a *LOG*-scale versus momentum transfer (Q), for both ambient (black) and cryogenic (red) temperature conditions obtained from ED patterns in (a) and (b), respectively.

Following these measurements on uncoated CNTs, we then examined the effect of cryogenic temperatures on beam damage effects in $AlO_x$ ALD-coated CNTs. We compare the results for ED patterns obtained for ALD-coated CNTs at cryogenic temperatures against ED performed at ambient temperature in Figure 6. We note a qualitative difference between the two different acquisition temperatures. Ambient temperature ED resulted in pronounced diffraction rings and texturing in Figure 6a. Cryogenic conditions resulted in diffuse smooth rings as depicted in Figure 6b, consistent with an amorphous material. In Figure 6c, we plot scattering intensity versus momentum transfer, Q, derived from the 2D diffraction patterns in Figure 6a and 6b. We observe in Figure 6c that the ED pattern obtained under ambient temperature conditions shows sharp peaks whereas the ED pattern obtained under cryogenic temperature conditions displays diffuse peaks, consistent with the 2D ED patterns shown in Figure 6a and 6b, respectively.

The diffuse smooth rings in the diffraction pattern observed under cryogenic temperature conditions in Figure 6b are in line with an amorphous material, as expected for ALD $AlO_x$.[16, 23, 64, 103, 104] The sharp features under ambient temperature diffraction arise from crystallization of the sample induced by beam damage (*vide infra*) as depicted in Figure 4b. The qualitative difference between ambient and cryogenic temperature diffraction data in Figure 6 shows that the beam damage effect is kinetically controlled at high beam energy (200 kV accelerating voltage) and can be minimized using cryogenic temperatures, in agreement with previous observations.[49] Under ambient temperature conditions, we observe an increase in diffraction intensities at peak locations (values of Q), consistent with both CNTs (~1.8 Å$^{-1}$)[68] and aluminum oxide (~4.2 Å$^{-1}$)[102] compared to diffraction intensities obtained when performed at cryogenic temperatures. This suggests that at ambient temperatures, both the CNTs and the ALD AlO$_x$ undergo substantial beam damage. After electron beam exposure under ambient temperature conditions, the atomic structure of the CNTs and AlO$_x$ are therefore expected to have changed and to no longer be representative of the materials present in the as-formed samples. This underscores the importance of mitigating beam damage effects in non-biological samples and the need for cryogenic temperatures to evaluate inorganic samples by TEM.[59] Interestingly, the diffraction intensity from the CNT is substantially lower for the AlO$_x$-coated CNT in Figure 6c than for the bare CNT in Figure 5c. This is surprising, and suggests that the ALD coating protects the underlying CNT substrate from electron beam damage. This protection effect is only observed at cryogenic temperature conditions, suggesting that the protection effect is kinetically mediated. The present data does not provide insight into the mechanism for protection, and so we refrain from speculation, but look forward to investigating this effect further. We emphasize that the samples measured in Figures 5 and 6 were not irradiated at high magnifications prior to the measurement, but experienced low magnification, low current beam for specimen localization, similar to "low dose mode" used in life sciences cryo-electron microscopy.[105, 106]

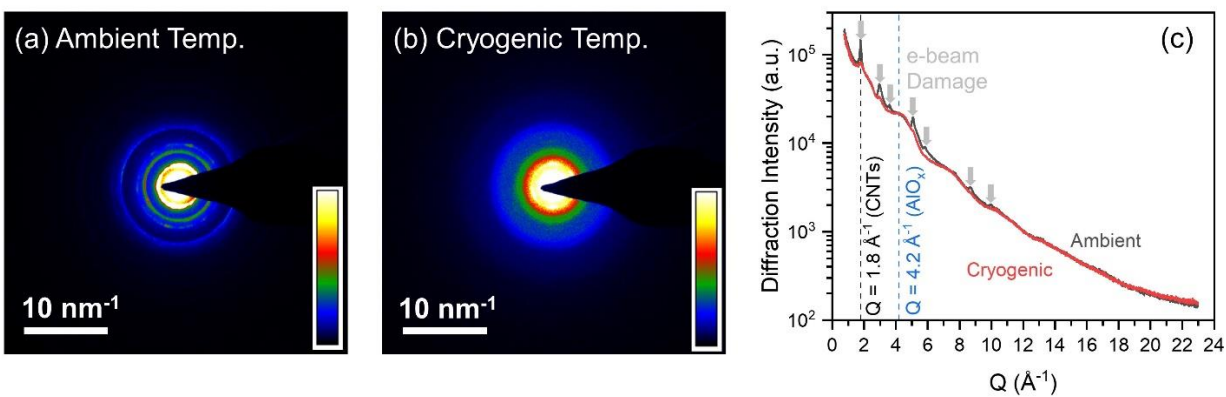

Figure 6. ED of ALD AlOx-coated CNTs deposited using 100 ALD cycles at 150°C under (a) ambient or (b) cryogenic temperature measurement conditions from 10 s electron diffraction exposure at a 200 kV accelerating voltage. The patterns have been colored using a heat map representing CCD electron counts with inset scalebars depicting relative diffraction intensity and reciprocal distance. (c) Total scattering intensity in a *LOG*-scale, for both ambient (black) and cryogenic (red) temperature conditions obtained from ED patterns in (a) and (b), respectively.

To further contrast the beam damage behavior under electron beam irradiation between cryogenic and ambient temperature conditions, we monitored changes in the ED pattern under both ambient and cryogenic conditions during 35 minutes of continuous electron beam irradiation as shown in Figures 7 and 8. In Figure 7 we show the diffraction intensity vs. Q under both ambient temperature and cryogenic temperature conditions at 0 min and 35 min of electron beam exposure. Under ambient temperature conditions in Figure 7a, we observe a distinct increase in the overall diffraction intensity (a vertical shift in the trace) after 35 min of electron beam exposure, and we note that some of the diffraction peaks (e.g. at Q=1.8, 3.0, 3.6, 5.1, and 7.7 Å$^{-1}$) become more pronounced after continued electron exposure. We note that these peak locations match the peak locations for the bare CNT substrate as shown in Figure 5c. We also note that, in general, an increase in background intensity could arise from the formation of carbonaceous material during imaging, but we did not observe the formation of any carbonaceous deposits on the surface. In contrast to ambient temperature conditions, under cryogenic temperature conditions in Figure 7b we observe minimal change in the diffraction pattern after 35 min of continuous electron beam exposure. Based on these data, we conclude that cryogenic temperatures reduce the rate of beam damage in the AlO$_x$-coated CNTs samples.

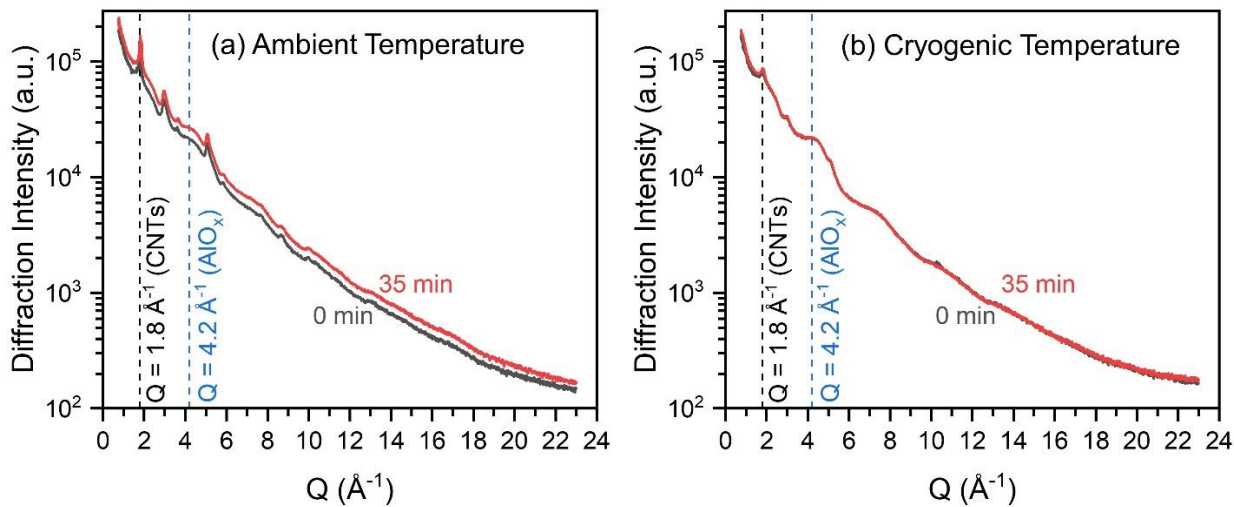

Figure 7. Total scattering intensity versus Q measured for ALD-grown aluminum oxide deposited on a CNT substrate using 100 ALD cycles at 150°C under extended electron beam exposure at (a) ambient temperature conditions and (b) cryogenic temperature conditions.

To more quantitatively evaluate the impact of beam damage under both ambient and cryogenic conditions, in Figure 8 we report changes in the peak area for peaks centered at Q values of 1.8 Å$^{-1}$ (integral over the range from 1.5 to 2.0 Å$^{-1}$) and 4.2 Å$^{-1}$ (integral over the range from 3.8 to 4.8 Å$^{-1}$). These peaks at Q=1.8 Å$^{-1}$ and 4.2 Å$^{-1}$ are associated with the CNTs and AlO$_x$, respectively, as described above. Peak areas reported in Figure 8 were normalized to the peak area observed under cryogenic conditions at a time of zero (or 10 s exposure). For the peak at Q=1.8 Å$^{-1}$ (associated with the CNT) we see an 18% increase in peak intensity after the initial 10 s of exposure (first diffraction measurement) at ambient temperature relative to a 10 s measurement at cryogenic temperature (first diffraction measurement). In addition to this rapid increase in diffraction intensity occurring during the first 10 s, we observe a monotonic increase in the diffraction intensity over the subsequent measurements up to 35 minutes. This proceeds at a steady state increase of ~1%/min from 14-35 minutes of electron beam exposure. These data indicate that under ambient temperature conditions, beam-induced crystallization occurs rapidly within the CNT during the first 10 s of electron beam exposure, and then slow to a moderate rate thereafter. We note that after 10 s of electron beam exposure under the diffraction imaging conditions used here (which are in line with typical diffraction imaging conditions used in materials science) the electron fluence is ~ 70 e$^-$/Å$^2$, which is significantly larger than the critical dose electron beam exposure of 10 e$^-$/Å$^2$ identified for organic materials in prior work.[107-109] Under cryogenic conditions, we see a negligible increase in the CNT diffraction intensity even after 35 minutes of electron beam exposure.

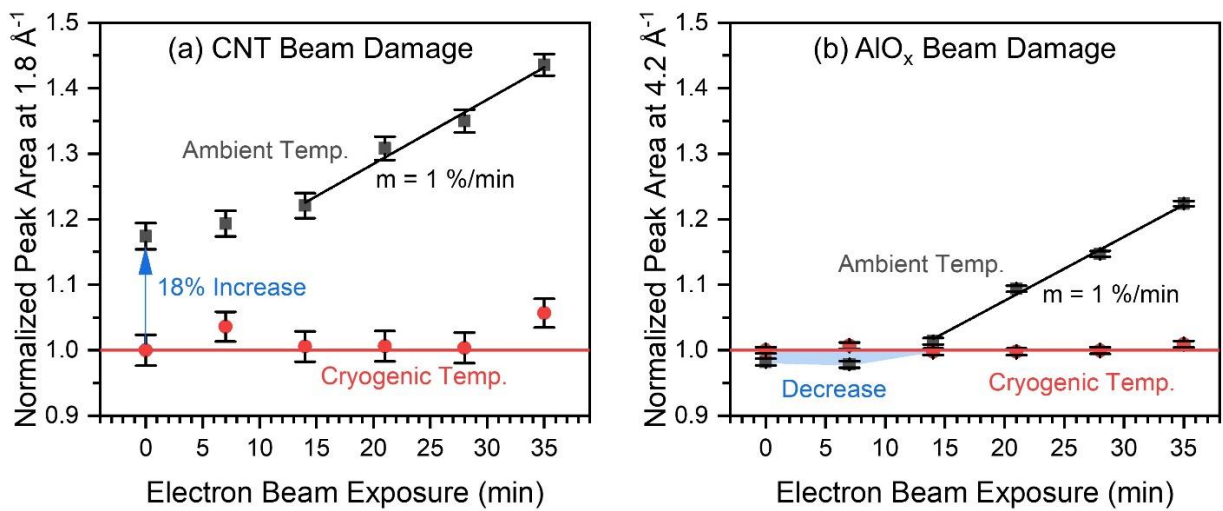

Figure 8. Plots of the peak area versus electron beam exposure time for an ALD AlO$_x$-coated CNT sample deposited using 100 ALD cycles at 150°C as measured under both ambient (black squares) and cryogenic (red circles) temperature conditions for peaks corresponding to (a) the dominant CNT feature at Q=1.8 Å$^{-1}$ and (b) the dominant AlO$_x$ feature Q=4.2 Å$^{-1}$. Peak areas were normalized to the peak area observed under cryogenic conditions at a time of zero (or 10 s exposure). Error bars in each panel are estimates of signal to noise obtained from cryogenic measurements.

At a first glance, we observe qualitatively similar beam damage behavior for the AlO$_x$ in Figure 8b as we observed for the CNT in Figure 8a. We observe that under ambient temperature conditions the peak area at Q=4.2 Å$^{-1}$ (corresponding to the AlO$_x$ structure) gradually increases at constant rate of increase of ~1% per minute of exposure. We also observe that cryogenic conditions bring this beam damage effect to a negligible rate. However, unlike the increase in peak area observed during the initial 10 s of electron beam exposure for the CNT peak, we observe a decrease in the peak area at ambient temperature (relative to cryogenic temperature conditions) during the initial 10 s of electron beam exposure for the AlO$_x$, which persists up to 14 minutes of electron beam exposure. This is surprising and seems to disagree with the depiction that beam damage forms crystallites in the AlO$_x$ as observed Figure 4b. We note that the beam damage in Figure 4b resulted from a high-intensity electron beam flux unlike the nanobeam conditions used to obtain the data in Figure 8. One possible explanation for the decrease peak area under ambient temperature conditions in Figure 8b up to 14 minutes of electron beam exposure is that structural variability is induced in the amorphous AlO$_x$ by the electron beam, followed by subsequent crystallization. This explanation is consistent with prior observations showing that beam damage can degrade crystalline materials and disrupt the crystalline order, leading to a decrease in Bragg intensities and amorphization.[51] We expect that this effect manifests here due the formation of charged defects under electron beam irradiation at ambient temperature conditions, which initially introduces more structural variability and reduces order in the material, then leading to crystallization at longer exposure times. We note that this effect is distinct from the crystallization effect described surrounding Figure 4b above, and underscores the complexity of beam damage effects which can take place in a given material.

We find that cryogenic temperature conditions dramatically slow beam damage in both the CNTs and AlO$_x$ in Figure 8a, and 8b, respectively. Under cryogenic temperature conditions, the normalized peak area for the CNT at Q=1.8 Å$^{-1}$ had an average value of 1.02 ± 0.02 over the 6 measurements from 0 to 35 minutes as reported in Figure 8a. Likewise, the normalized peak area for the $AlO_x$ at Q=4.2 Å$^{-1}$ had an average value of 1.002 ± 0.005 over the 6 measurements from 0 to 35 minutes reported in Figure 8b under cryogenic temperature conditions. This is an interesting result because it suggests that the use of cryogenic temperatures prevents beam damage in both the carbonaceous CNT and the inorganic $AlO_x$ layers within the ALD-coated CNT sample. The local order of both materials is preserved for electron beam exposure durations up to 35 minutes under cryogenic temperatures. For the ~7 e$^-$/Å$^2$/s electron flux calculated under these beam conditions, 35 minutes of exposure corresponds to 14,700 e$^-$/Å$^2$. This suggests that beam damage alters the structure of inorganic and carbonaceous materials under ambient temperature, but cryogenic temperatures dramatically limit the rate of beam damage and enable the prolonged study of inorganic and carbonacoues materials under electron beam exposures up to 35 minutes.

After establishing that cryogenic temperatures minimize beam damage during ED measurements, we then proceeded to probe the local structure of the amorphous $AlO_x$ grown by ALD using ED and PDF analysis using the processing steps as outlined in Figure 2 above. Figure 9a shows ED pattern obtained under cryogenic conditions at a 200 kV acceleration voltage for the sample shown in Figure 3a, which was deposited at 200°C. The ED diffraction pattern in Figure 9a shows no evidence of crystallization from beam-damage. The PDF, or G(r), generated from the ED pattern in Figure 9a is plotted in Figure 9b along with a PDF generated from HE-XRD measurements on ALD AlOx from previous work.[23] HE-XRD data was collected on ALD $AlO_x$ films grown on the same CNT-OH substrate and using the same ALD precursors employed here. The final ALD $AlO_x$ film thickness was also approximately the same between the samples (11 nm for the ED sample and 14 nm for the HE-XRD sample). As such, the ED-PDF data and HE-XRD PDF in Figure 9b are expected to be similar. Indeed, both have dominant peaks centered at ~1.8 Å which correspond to the first coordination shell (Al-O bonds). The peak center for this feature (representing the Al-O bond length) is located at 1.829 Å from the X-ray characterization vs. 1.823 Å for the ED PDF measurement. Both samples also exhibit broad peaks centered at ~3 Å and ~4.5 Å, which are assigned to overlapping features from the second and third coordination shells, respectively, of multiple Al⋯Al, Al⋯O, and O⋯O atomic pairs in the amorphous structure, and both PDF patterns exhibit minimal structural coherence at pair distances > 5 Å, consistent with the amorphous character of ALD-grown $AlO_x$.[16, 23, 64, 103, 104] We note that the $Q_{max}$ used for PDF calculation from ED measurements was 22 Å$^{-1}$, whereas $Q_{max}$ from HE-XRD measurements was 30 Å$^{-1}$. The higher maximum $Q_{max}$ used for HE-XRD allows for resolving finer structural details.

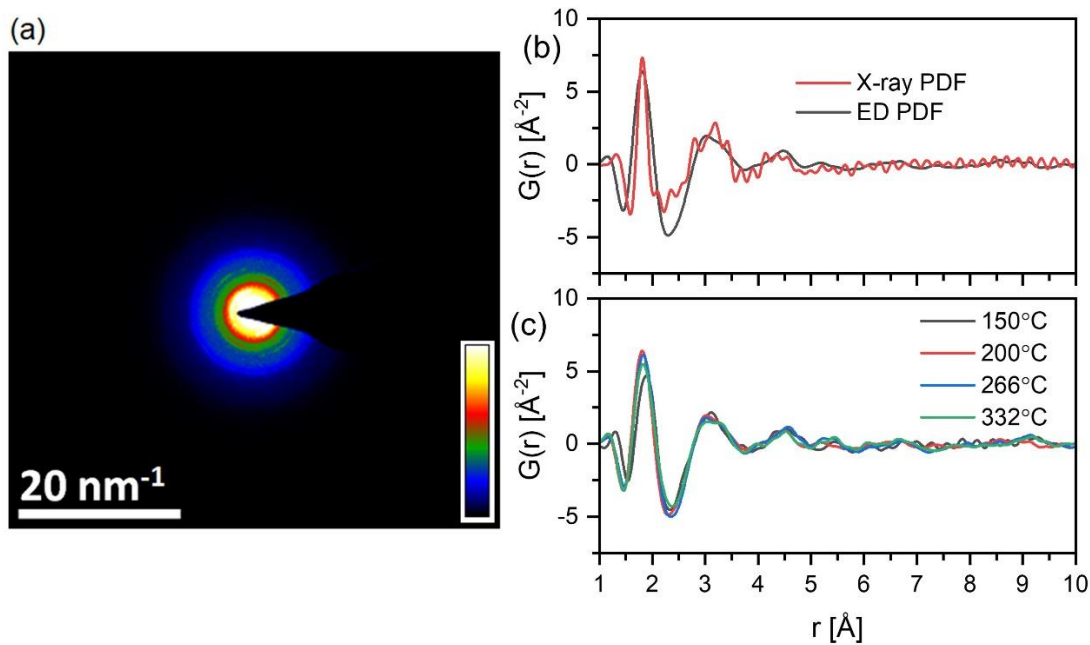

Figure 9 (a) ED of AlOx on CNTs substrate under cryogenic conditions with inset scalebars depicting relative diffraction intensity and reciprocal distance. (b) PDF analyses for both TEM-based cryo-ePDF (black) and HE-XRD PDF (red). (c) cryo-ePDF curves generated for four samples with ALD AlOx deposited at different deposition temperatures (150°C is black, 200°C is red, 266°C is blue, and blue, and 332 °C is green), each prepared using 100 ALD cycles.

After benchmarking the ED PDF data against HE-XRD data, we then expanded on this previous HE-XRD work by studying the atomic structure of ALD AlO$_x$ deposited at varying temperatures using the cryo-ePDF conditions reported here. In Figure 9c we report cryo-ePDF traces for ALD AlO$_x$ deposition temperatures from 150-332 °C. The cryo-ePDF data are similar for the varying deposition temperatures, but we do observe variations in the cryo-ePDF data. These variations indicate that the AlO$_x$ structure changes with deposition temperature. For example, as the deposition temperature increases from 150°C to 200°C, the peak centered at a pair distance of ~1.8 Å (corresponding to Al-O bonds) increases in intensity and shifts to a lower pair distance. The increase in area under the curve at this pair distance indicates that the Al-O coordination number is increasing, whereas the peak shift to lower pair distance indicates that the average Al-O bond length is decreasing. As the temperature increases from 200°C to 332°C, there is a slight decrease in area under the curve for the peak centered at a pair distance of ~1.8 Å, indicating a decrease in Al-O coordination number, and no obvious shift in the peak center, indicating no change in the Al-O bond length.

We also note variations in the PDF at larger pair distances (e.g. at ~ 3 Å) as the deposition temperature changes. However, as discussed in our previous work,[23] it is challenging to qualitatively interpret these variations in PDF data at higher pair distances by visual inspection because the changes may arise from varying sources. By employing stochastic structural modeling (i.e. RMC modeling) we are able to establish atomic structures which are consistent with experimental observations over the full range of PDF data. These models were then statistically analyzed to understand how growth temperature influences a range of atomic structure parameters. Figure 10a shows PDF analysis measured from ED compared with a computed PDF from an RMC simulation. The measured data are in a good agreement with the RMC model data. Figure 10b shows a stochastic structural model derived from the RMC fitting. Similar structural models were established for each of the growth temperatures, and were analyzed to yield the structural metrics reported in Figure 10c-f.

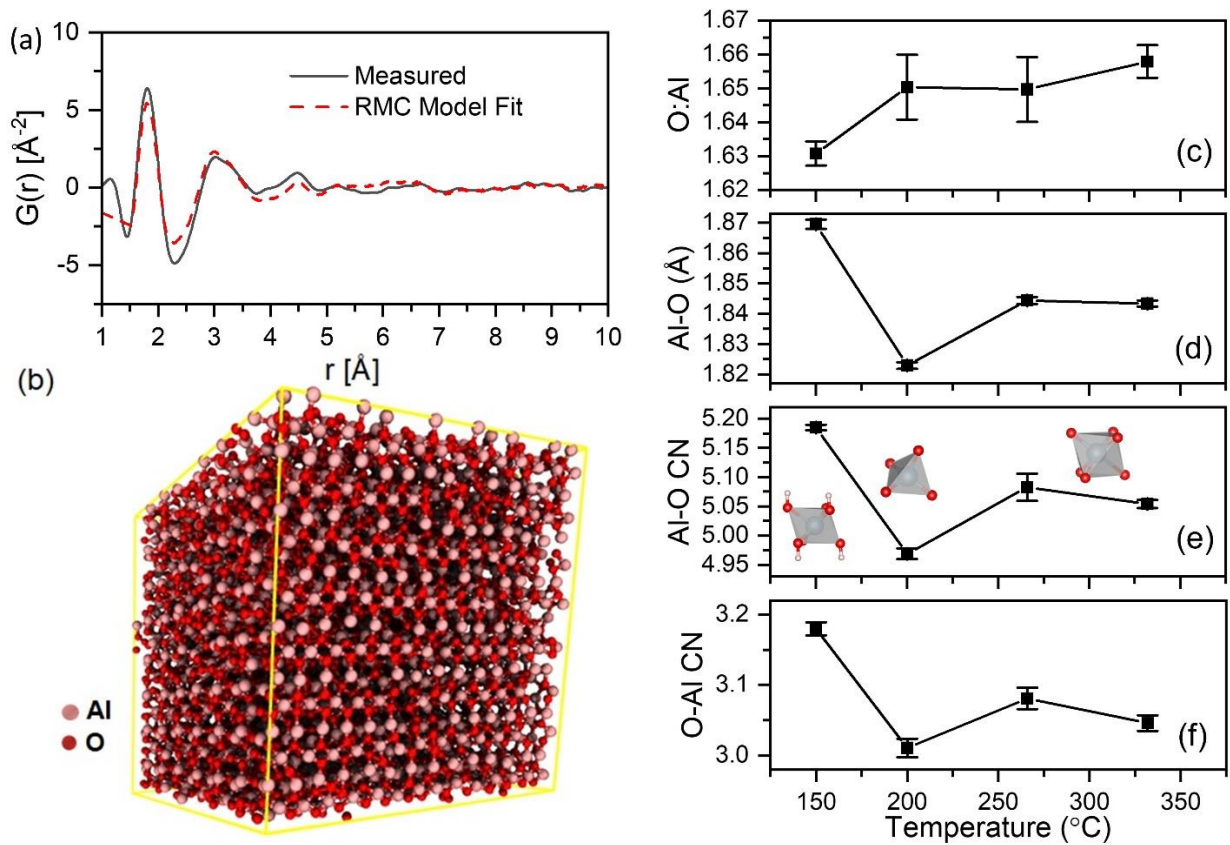

Figure 10. Results of RMC modeling including (a) experimental PDF at 200°C deposition temperature versus the PDF generated from RMC fitting and (b) a resulting atomic structural model from the RMC fit at 200°C, and the structural metrics derived from RMC structural fitting versus temperature including (c) stoichiometric ratio of O:Al, (d) Al-O bond lengths, (e) Al-O CN and (f) O-Al CN. The values reported in (c-f) are average values from at least three RMC modeling runs performed for each sample, where the error bars represent the standard deviation in modeling results.

We observe the highest values of Al-O bond length, Al-O CN and O-Al CN, and the lowest value of O:Al ratio at a deposition temperature of 150°C relative to 200-332°C. These outlying values at 150°C can largely be explained by differences in hydrogen content at this lower deposition temperature. We note that prior work found that the hydrogen content in ALD AlO$_x$ increases with decreasing growth temperature from <2% at 300°C to ~5% at 150°C.[14] The higher hydrogen content at lower growth temperatures can be attributed to the presence of hydroxyls within the AlO$_x$ structure. This is in line with the data in Figure 10c-f, where the reported metrics are consistent with the presence of some aluminum hydroxide at a 150°C deposition temperature. We note that the α-Al(OH)$_3$ structure contains exclusively octahedral AlO$_6$ groups (Al-O CN = 6) with Al-O bonds ranging from 1.84-1.95 Å.[110] For comparison, Al-O bond lengths within θ-Al$_2$O$_3$ (with no hydroxyls present) are in the range of 1.7-1.79 Å for $AlO_4$ tetrahedra and 1.99-2.10 Å for $AlO_6$ octahedra.[111] Considering these reference points, the higher values of Al-O bond length and Al-O CN at 150°C are consistent with α-Al(OH)$_3$ and suggest the presence of excess hydroxyls at this growth temperature. However, the values of O-Al CN and O:Al we observe at 150 °C relative to the other growth temperatures seem to be inconsistent with this picture. The O:Al ratio is expected to be 1.5 for θ-$Al_2O_3$ and 3 for α-Al(OH)$_3$, but we observe a smaller value of O:Al at a growth temperature of 150 °C relative to higher growth temperatures. Likewise, the O-Al CN is expected to be 3.5 for θ-$Al_2O_3$ and 2 for α-Al(OH)$_3$, but we observe an increased value at lower growth temperatures.

To reconcile these seemingly contradictory trends in the O-Al CN and the O:Al ratio against the picture of changing H content in the as-grown films described above, we consider how the growth temperature may influence the growth mode of TMA/$H_2O$ ALD. The TMA ALD half reaction requires surface hydroxyls, and the equilibrium surface hydroxyl density is expected to decrease at elevated growth temperature.[12, 15, 64, 80-84] This decrease in hydroxyl density with increasing temperature arises from recombinative desorption of *OH groups (* indicates a surface-bound species) to release $H_2O$ via the reaction 2*AlOH → *Al-O-Al* + $H_2O$.[64] The decrease in surface hydroxyl coverage at elevated temperature is significant, with an expected ~40% reduction in surface *OH density on the growth surface upon increasing the growth temperature from 150°C to 300°C.[80, 81] The decrease in the number of *OH growth sites at elevated growth temperatures is expected to (a) introduce a preference for TMA to react via single-reaction sites (Al(CH$_3$)$_3$ + *OH → *OAl(CH$_3$)$_2$) over dual reaction sites (Al(CH$_3$)$_3$ + 2*OH → *O$_2$AlCH$_3$) during the TMA ALD half reaction, and (b) lead to the incorporation of unreacted Al-O-Al groups within the bulk AlOx structure. We believe the combination of these effects would lead to a higher O-content and lower O-Al CN values at elevated growth temperatures, consistent with the observations in Figure 10c and Figure 10f, respectively.

In Figure 11 we present a schematic depiction of our interpretation of the influence of growth temperature on the growth surface based on known changes in surface hydroxyl coverage in connection with the experimental local structure data reported in Figure 10. We note that these 1D surface representations convey a qualitative picture to interpret Figure 10c-f, and do not capture the complete growth picture during deposition on a 2D area. Our interpretation of the data in Figure 10 is that as the growth temperature increases, the available functional groups for the TMA surface reaction change, ultimately leading to differences in the final molecular structure. Specifically, we envision the end-formation of a blend of $AlO_x(OH)_y$ octahedra, $AlO_4$ tetrahedra, and $AlO_6$ octahedra at each growth temperature, where the surface hydroxyl density shifts with growth temperature and alters the relative concentrations of each of these constituent structural components.

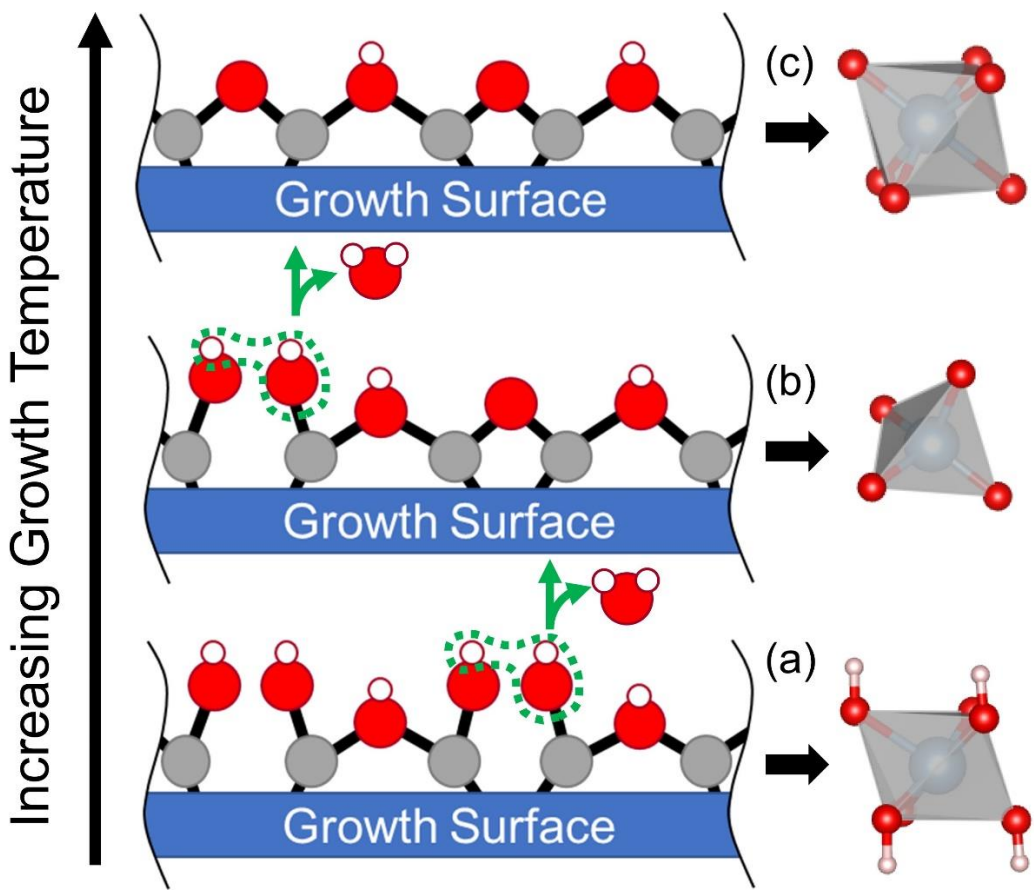

Figure 11. Qualitative depiction of the influence of growth temperature on the growth surface during TMA/$H_2O$ ALD, and the resulting impact on structural features present in the ALD films, where the grey, red, and white spheres represent Al, O, and H, respectively.

At a growth temperature of 150°C, we expect a larger amount of surface hydroxyl groups during ALD growth will lead to the formation of more $AlO_x(OH)_y$ octahedra within the final structure as depicted in Figure 11a, in line with the longer Al-O bond length and higher Al-O CN at this growth temperature in Figures 10c and 10e, respectively. At a growth temperature of 200°C, the surface hydroxyl density during growth is lower than for a growth temperature of 150°C, and the structure composition shifts to less $AlO_x(OH)_y$ octahedra and more $AlO_4$ tetrahedra as depicted in Figure 11b, in line with the shorter Al-O bond length and lower Al-O CN in Figures 10c and 10e, respectively. As the growth temperature increases above 200°C and the number of surface hydroxyls available during growth continues to decrease, we expect that Al metal centers will be more sparsely dispersed on the growth surface after TMA reaction, allowing the Al metal centers to coordinate with surrounding surface oxygen atoms, increasing the final Al-O CN as observed in Figure 10e and leading to more $AlO_6$ octahedra as depicted in Figure 11c. This overall trend in transitioning from $AlO_x(OH)_y$ → $AlO_4$ → $AlO_6$ with increasing growth temperature is consistent with observations that the $AlO_x$ structure densifies at increasing growth temperature.[15] At a growth temperature of 150 °C, OH groups trapped within the $AlO_x$ structure limit the densification of the atomic structure. We note that while the densification of the $AlO_x$ layer at elevated temperatures might be expected to yield continually shorter average Al-O bond length in Figure 10d with increasing temperature, we suggest that more $AlO_6$ units form (as indicated by the Al-O CN) as the growth temperature increases above 200°C, which is consistent with an increased bond length (Al-O bond lengths are 1.7-1.79 Å for $AlO_4$ tetrahedra and 1.99-2.10 Å for $AlO_6$ octahedra[111]). We acknowledge that the proposed picture we describe will require additional data and follow-on studies to examine further, and that other variables such as purge time and precursor exposure are also expected to influence the structural metrics discussed here.

Finally, we note that although most of the metrics reported in Figure 10c-f are in reasonable agreement with our previous report examining the structure of ALD $AlO_x$ *in operando* using HE-XRD, the Al-O CN values measured in our previous work were consistently < 5,[23] but we observe Al-O CN values as high as ~5.2 here. We suspect that a CVD-component may have contributed to growth under the growth conditions used in our prior work, where the reactor geometry required for *in operando* HE-XRD limited the efficiency of the purge.[23] This is consistent with the higher thickness observed on the CNTs in the *in operando* experiments (14 nm over 50 ALD cycles vs. 10 nm over 100 ALD cycles here). We propose that this CVD-component to growth may have contributed to the difference in Al-O CN values between the two studies. Specifically, the tetrahedral structure of TMA dimers[112] may act to template tetrahedral $AlO_4$ sites within $AlO_x$, giving rise to lower average coordination numbers in the ALD $AlO_x$ during TMA/$H_2O$ growth with a CVD component.

**Conclusions:**

Broadly, this work addresses a key hurdle to employing ePDF analysis for atomic structure measurements of amorphous inorganic phases within a TEM by demonstrating that cryogenic temperatures, or cryo-ePDF, can dramatically slow the rate of beam damage during electron diffraction. Although cryogenic temperatures have been broadly applied to enable TEM characterization of small molecules and biomolecules,[56-58] and are increasingly recognized for their benefits in materials science TEM studies,[59] to the authors' knowledge, this work represents the first study of cryogenic temperatures during ePDF characterization. This work suggests that cryo-ePDF within a TEM may be viable for routine atomic structure characterization of amorphous inorganic phases. These cryo-ePDF measurements could act to complement synchrotron HE-XRD/PDF measurements and can be performed on a more rapid timescale using TEM instrumentation available on many university campuses. The use of cryo-ePDF characterization also promises to enable the elucidation of more localized structural information versus X-ray PDF characterization, where the electron beam within a TEM can be focused to sub-micron, even nanometer scale areas[29-31] to characterize atomic structure with unprecedented spatial resolution. Localizing the electron beam in this way also promises to enable the study of amorphous coatings on crystalline substrates by cryo-ePDF, where the electron beam can be focused to avoid the crystalline substrate. Future work is needed to pursue these exiting paths and examine whether cryogenic temperatures sufficiently mitigate beam damage effects in other amorphous and nanoscale materials (e.g. other inorganic materials, polymers, metal nanoparticles) to enable ePDF characterization for a broader range of materials.

Within the field of ALD, where amorphous inorganic materials are ubiquitous, the ability to overcome beam damage using cryo-ePDF for structural characterization demonstrated in this work promises help to close the loop on process-structure-property understanding and enable researchers to innovate ALD coatings to address technological needs more rapidly, and with a higher success rate. Especially the promise of focusing the electron beam to a small area using ED within a TEM[29-31] to characterize the structure of ALD coating will help tremendously in understanding the atomic structure of ultrathin ALD coatings, and, for example, help to reveal how the substrate impacts the atomic structure of the ALD films. The connection between growth temperature, surface hydroxyl density, and resulting structural composition established in this work for ALD-grown aluminum oxide is expected to influence experimental design of ALD-grown coatings, and is a testament to the promise the cryo-ePDF technique holds for improving understanding of amorphous ALD-grown materials in particular.

## ■ ASSOCIATED CONTENT

## ■ AUTHOR INFORMATION

Corresponding Authors: Tommi A. White and Matthias J. Young

* Email: : whiteto@missouri.edu or matthias.young@missouri.edu
Ahmed M. Jasim: https://orcid.org/0000-0002-7649-3927

Xiaoqing He:

Yangchuan Xing: https://orcid.org/0000-0002-5985-3222

Tommi A. White: https://orcid.org/0000-0001-7283-3755

Matthias J. Young: https://orcid.org/0000-0001-7384-4333

### Author Contributions

All authors have given approval to the final version of the manuscript.

### Notes

The authors declare no competing financial interest.

## ■ Acknowledgement

AMJ and MJY thank the Electron Microscopy Core (EMC) and the Office of Research at the University of Missouri for financial support through the Excellence in Electron Microscopy award. MJY acknowledges support from faculty start-up funds from the University of Missouri. We thank the Prof. Nicholas Bedford at the University of New South Wales for useful discussion regarding PDF analysis.

For Table of Contents Only:

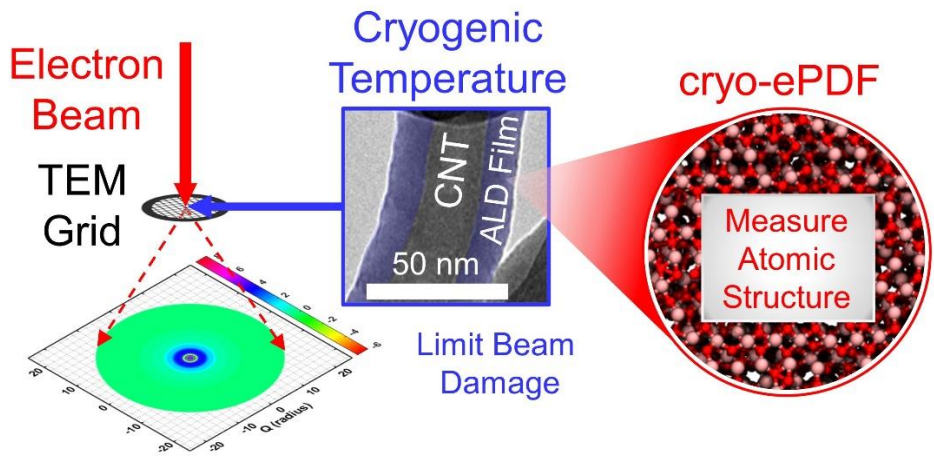